# Light-modulated exchange bias in multiferroic heterostructures

*Huan Tan[1,2]\*, Zheng Ma[1], Cynthia Bou Karroum[3], Matthieu Liparo[3], Jean-Philippe Jay[3], David Spenato[3], David T. Dekadjevi[3], Luís Martínez Armesto[1,2], Alberto Quintana[1,2], Jordi Sort[1,2,4]\**

[1]Departament de Física, Universitat Autònoma de Barcelona, E-08193 Cerdanyola del Vallès, Spain.

[2]Catalan Institute of Nanoscience and Nanotechnology (ICN2), CSIC and BIST, Campus UAB, Bellaterra, 08193 Barcelona, Spain

[3]Univ. Brest, Laboratoire d'Optique et de Magnétisme (OPTIMAG), UR 938, 29200 Brest, France.

[4]Institució Catalana de Recerca i Estudis Avançats (ICREA), Pg. Lluís Companys 23, E-08010 Barcelona, Spain

E -mail: Huan.Tan@uab.cat (H. Tan), Jordi.Sort@uab.cat (J. Sort)

## Abstract

Magnetic straintronics, the strain-mediated control of magnetic anisotropy, has emerged as a key direction for next-generation energy-efficient technologies. In multiferroic heterostructures, magnetoelectric coupling is typically achieved by applying an electric field on a ferroelectric phase, inducing strain through the converse piezoelectric effect, which is then transferred to the adjacent ferromagnetic phase. As an alternative, strain can be remotely modulated through the photostrictive effect induced by light. While light-driven control of magnetic anisotropy has been explored, optical modulation of more complex phenomena such as exchange bias remains largely unaddressed. Here, we demonstrate significant light-induced modulation of exchange bias and magnetization switching at room temperature in a Pb(Mg$_{1/3}$Nb$_{2/3}$)O$_3$-Pb(Zr,Ti)O$_3$ (PMN-PZT)/Fe$_{80}$Ga$_{20}$(FeGa)/Ir$_{20}$Mn$_{80}$(IrMn) multiferroic heterostructure, driven by visible-light-photostriction. The magnetization state correlates with the light intensity, enabling multi-level states with light power densities as low as 0.1 W cm$^{-2}$. These findings suggest a promising route toward low-power, multistate, and wireless opto-magnetic memory applications.

# 1. Introduction

The growing demand for energy-efficient, high-density, and non-volatile memory technologies has spurred intense interest in multiferroic heterostructures, [1-3] where electric, magnetic, and elastic orders coexist and are inter-coupled. Several methods have been proposed to control magnetization in these materials, such as exchange coupling,[4] charge-modulation,[5] and strain-induced magnetoelectric interaction between the ferromagnetic and ferroelectric counterparts.[6] Among these various magnetoelectric mechanisms, magnetic straintronics, i.e., the strain-mediated modulation of magnetic properties, has emerged as a particularly promising strategy for realizing low-power, fast-switching spintronic devices.[7] In such systems, strain is typically generated via the converse piezoelectric effect in a ferroelectric (FE) substrate and transferred to an adjacent ferromagnetic (FM) layer, enabling voltage control of magnetization.

Strain-mediated magnetoelectric (ME) coupling has been extensively studied in multiferroic heterostructures comprising high-performance piezoelectric single-crystal substrates such as $Pb(Mg_{1/3}Nb_{2/3})O_3$–$PbTiO_3$ (PMN-PT) and $Pb(Zn_{1/3}Nb_{2/3})O_3$–$PbTiO_3$ (PZN-PT) and $Pb(Mg_{1/3}Nb_{2/3})O_3$-$Pb(Zr,Ti)O_3$ (PMN-PZT).[8-11] These materials offer exceptional piezoelectric coefficients and electromechanical coupling, enabling efficient strain transfer and robust magnetoelectric interactions. For instance, field-induced anisotropy fields up to 3500 Oe and ME coefficients as high as 580 Oe cm $kV^{-1}$ have been demonstrated in Terfenol-D/PZN-PT systems.[12] Electric-field-controlled magnetization reorientation has also been reported in Ni/PMN-PT,[13] FeGa/PMN-PZT[14] and FeGaB/PZN-PT (001) [15] heterostructures. However, conventional electric-field-based control of ferroelectric polarization suffers from limitations such as high operational voltages, dielectric breakdown, and material fatigue over repeated cycling. [16] These drawbacks have motivated the search for alternative, non-contact approaches for the strain generation. One such method involves photostriction, where strain is induced in a photoresponsive ferroelectric material under light illumination.[17] This approach offers several advantages, including remote operation, reduced circuit complexity, and enhanced endurance compared to electric-field cycled magnetoelectric composites.[18]

The phenomenon of photostriction was first reported by Tatsuzaki et al. in 1960 in SbSI single crystals,[19] where strain was shown to arise from the combination of the bulk photovoltaic effect (BPVE) and the converse piezoelectric effect. Since then, various photostrictive ferroelectrics and thin films, such as PLZT,[20, 21] $BaTiO_3$(BTO) ,[22] and $BiFeO_3$ (BFO),[23] have been explored for optically induced strain. Recent advances have included the role of charged ferroelectric domain walls in the photostrictive effect at the nanoscale in BTO.[24] Moreover, significant changes in magnetic coercivity, up to 75%, have been achieved in BFO/Ni [25] heterostructures under visible light excitation, highlighting the potential of photostriction for control of magnetism. Furthermore, light driven manipulation of axial anisotropy has been reported in FeGa/PMN-PZT[26] and FeGaLa/PMN-PZT[27] using visible laser. One of the bottlenecks in conventional photostrictive systems, however, is their inherently slow response time, often limited to tenths of a second or longer due to the gradual buildup of photovoltage. [28] Recent studies have overcome this limitation in PMN-PT crystals, where rapid and efficient photostrictive effects have been achieved through localized microscale fast BPVE-converse piezoelectric interactions, even without the need to polarize the ferroelectric material overall.[29] These results open new opportunities for fast, energy-efficient, and wireless control of magnetization via optical stimuli.

Despite these advances, the use of light to control more complex interfacial magnetic phenomena, such as *exchange bias*, remains largely unexplored. Exchange bias, which arises from interfacial coupling between an antiferromagnet (AFM) and a ferromagnet (FM),[30, 31] plays a central role in stabilizing magnetic states in spin valves, sensors, and memory elements. Traditionally, tuning exchange bias requires thermal annealing or external magnetic field cycling, which hinders its integration into low-power and reconfigurable devices. A wireless, light-based method to modulate exchange bias at room temperature would thus represent a significant technological breakthrough.

Ultrafast optical excitations of AFM/FM bilayers have been used to probe the interfacial interaction between the AFM and FM layers, and interesting fast magnetization dynamics have already been observed.[32-34] Notably, reversible, electric-field-free switching of exchange bias has been demonstrated in Fe/Pt heterostructures using femtosecond laser pulses via ultrafast photothermal modulation of interfacial spin coupling.[35] Meanwhile, helicity of circularly polarized light can also manipulate the exchange bias.[36, 37] Nevertheless, in all these works, laser-induced heating has been reported, which results in high energy consumption. Moreover, heating can trigger unstable switching of adjacent magnetic domains, thereby limiting storage density and stability. In contrast, soft X-ray illumination can reversibly switch exchange bias and create zero-field skyrmions in Pt/Co/IrMn multilayers through non-thermal, photon-induced AFM spin reconfiguration.[38] Recently, visible light has been shown to deterministically and reversibly switch magnetization in synthetic antiferromagnets via photoelectron-induced enhancement of the Ruderman–Kittel–Kasuya–Yosida (RKKY) interaction without significant heating.[39] At low temperatures (25 K), weak red-light irradiation (~1 mW cm$^{-2}$) has been reported to reduce the exchange bias field in $BiFeO_3/La_{2/3}Sr_{1/3}MnO_3$ (BFO/LSMO) heterostructures due to photo-carrier injection from the $SrTiO_3$ substrate.[40] Besides these approaches, ultraviolet light was reported to empirically tune exchange bias in $Co_{90}Fe_{10}/BiFeO_3$, although no clear explanations of the underlying mechanisms governing this effect were provided.[41] Therefore, exploring non-thermal optical control of exchange bias at room temperature is of great interest, especially in systems employing metallic antiferromagnets commonly used in spin valves and magnetic tunnel junctions– an essential step toward low-energy, scalable spintronic technologies.

In this work, we demonstrate a non-volatile light-induced modulation of exchange bias in a ferroelectric/ferromagnetic/antiferromagnetic (FE/FM/AFM) multiferroic heterostructure, without the need for applied electric fields or magnetic field cooling. We employ a (011)-oriented PMN-PZT single-crystal substrate, chosen for its large piezoelectric coefficients and a relatively low bandgap (3.03 eV),[26] enabling efficient visible-light-induced photostrictive strain. This strain is transferred to an $Fe_{80}Ga_{20}$ ferromagnetic layer interfaced with an $Ir_{20}Mn_{80}$ antiferromagnetic layer, leading to a modulation of magnetic anisotropy and interfacial exchange bias coupling. Furthermore, we observe tunable, multi-level exchange bias fields and magnetization states as a function of light intensity, with significant modulation achieved at power densities as low as 0.1 W cm$^{-2}$ under ambient conditions. These findings establish a novel route for non-thermal light-controlled manipulation of interfacial magnetism, expanding the functionality of strain-mediated magnetoelectric systems and paving the way for wireless, multistate, opto-magnetic memory technologies.

## 2. Results and discussion

Figure 1a, b show a schematic diagram and a cross-sectional elemental mapping (via energy-dispersive X-ray spectroscopy, EDX, using transmission electron microscopy, TEM) of the PMN-PZT/Ta/FeGa/IrMn/Ta heterostructure. Here the PMN-PZT is a (011)-oriented single crystal. Individual maps of the different elements comprised in the stack, together with a compositional line profile across the film and a high-resolution TEM image (revealing the polycrystalline nature of FeGa/IrMn) are shown in the Supporting Information (Figure S1). The multilayer stack was deposited by sputtering under an in-situ magnetic field, $H_{sputtering}$, of approximately 2000 Oe, applied using a permanent magnet positioned near the sample. Two samples were grown, in which $H_{sputtering}$ was aligned along the [0−11] and the [100] crystallographic in-plane directions of the PMN-PZT crystal, respectively, to exploit the anisotropic strain effect. The presence of $H_{sputtering}$ establishes an in-plane magnetic easy axis in the FeGa film during the growth, and it enables unidirectional exchange coupling between the IrMn and FeGa layers, without requiring a field cooling (FC) process. In the coordinate system shown in Figure 1a, $\theta$ represents the angle between $H_{sputtering}$ (*i.e.*, the direction along which exchange bias is induced) and the magnetic field applied while recording the subsequent hysteresis loops. As shown in Figure 1c, a significant shift of the magnetization hysteresis (*M-H*) loop at $\theta = 0°$ is observed in the pristine as-prepared state, which indicates a unidirectional exchange coupling effect between the ferromagnetic and antiferromagnetic layers is induced during the deposition. The loops shift, $H_{EB}$, is calculated as $H_{EB} = |(H_{C1} + H_{C2})/2|$, where $H_{c1}$ and $H_{c2}$ are the coercivity values for the descending and ascending branches of the *M-H* loops, respectively. A training effect commonly occurs in exchange bias systems,[42] as the antiferromagnetic domain structure reconfigures during magnetization reversal of the ferromagnetic layer under repeated magnetic field cycling. To eliminate the influence of the training effect, several consecutive hysteresis loops (10 in total) were recorded prior to illumination (as shown in Figure S2a), until $H_{EB}$ stabilized (as-prepared state). The same repeated magnetic field cycling was applied to the other sample (i.e., with $H_{sputtering}$ ∥ [100]), whose magnetic easy axis is parallel to the [100] direction of PMN-PZT single crystal, as shown in Figure S2b.

To investigate the effects of light illumination, a visible blue laser (405 nm wavelength, corresponding to a photon energy of 3.06 eV) was directed onto the backside of the PMN-PZT (011) single crystal (as illustrated Figure 1a). The influence of light on the magnetic properties was evaluated by comparing measurements performed in the dark ("Light Off") and under illumination ("Light On"). Under illumination, a notable reduction of the exchange bias field was observed along the [0−11] direction, from 252.9 Oe to 231.5 Oe (Figure 1c), which was accompanied with virtually no change in the magnitude of coercivity ($H_C$) and remanent magnetization ($M_R$). After illumination, the exchange bias field tends to recover, with time, back to the as-prepared state (as displayed in Figure 1c, i.e., "Light Off 1st"). There is a slight difference between the as-prepared state and the "Light Off 1st" state, evidencing that a strain relaxation effect exists in the ferroelectric PMN-PZT single crystal. 2nd illumination is applied subsequently, resulting a similarly pronounced modulation in exchange bias field, with no observable change in $H_C$. Once the illumination is removed ("Light Off 2nd"), the M–H loop returns to its initial state. The sample was precycled to eliminate any influence from the training effect. The repeatable response under light illumination indicates that the modification of the exchange bias field is non-thermal and stable, arising from the photovoltaic effect in the PMN-PZT single crystal, combined with strain-mediated magnetoelectric coupling with the FeGa/IrMn bilayer. If the observed modulation were dominated by thermal effects, one would expect a progressive decrease in $H_{EB}$ with repeated light cycling, due to irreversible reconfiguration of the interfacial spins, but this is not observed experimentally.

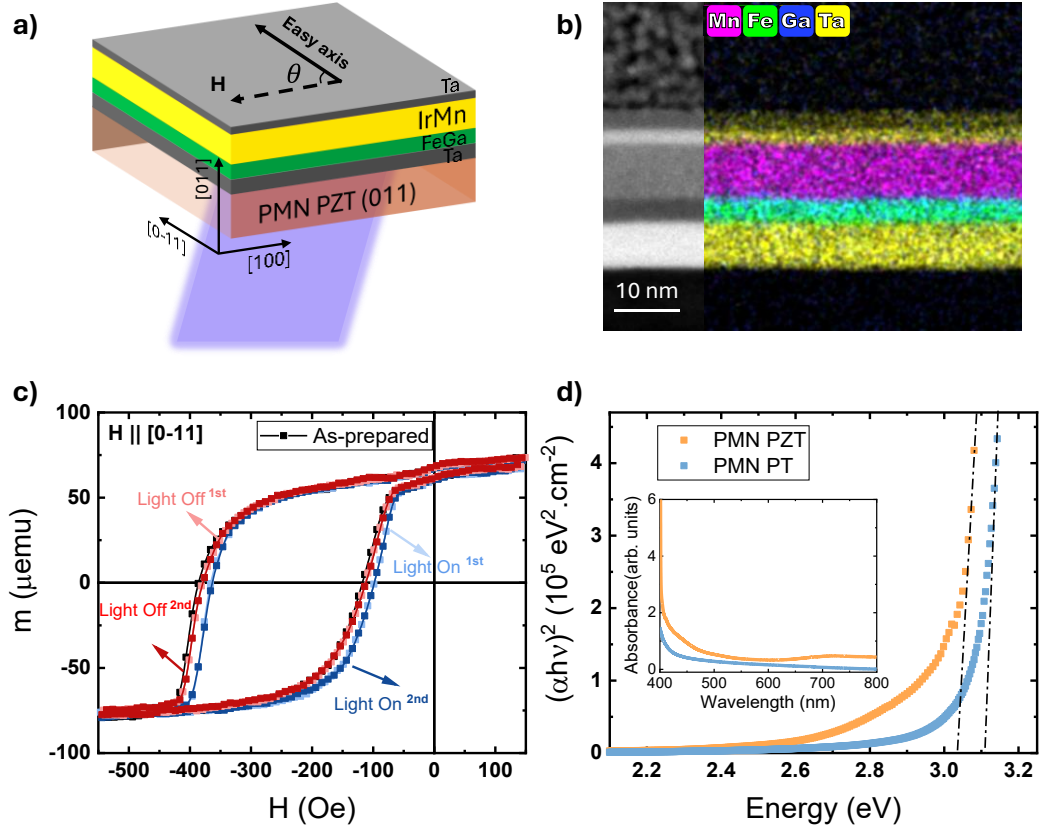

**Figure 1. a)** Schematic structure of the Ta (10 nm)/Fe$_{80}$Ga$_{20}$ (5 nm)/Ir$_{20}$Mn$_{80}$ (10 nm)/Ta (5 nm) on (011)-cut PMN-PZT heterostructure and experimental geometry. **b)** Cross-sectional elemental mapping of the multilayer stack (right), with the corresponding TEM image (left). **c)** Magnetic hysteresis loops along the magnetic easy axis in $H_{sputtering}$ ∥ [0−11] sample in dark and under illumination. 2$^{nd}$ illumination is applied subsequently after the 1$^{st}$ illumination. **d)** $(ahv)^2$ dependence on photoenergy, $hv$, for PMN-PZT and PMN-PT (011)-cut single crystals. The inset shows optical absorption spectra of the two single crystals as a function of the wavelength.

To confirm the photostrictive effect in PMN-PZT, the same AFM/FM heterostructure was also deposited on Si (100) and on Pb(Mg$_{1/3}$Nb$_{2/3}$)O$_3$-PbTiO$_3$ (PMN-PT) (011) substrates. The FeGa/IrMn films were deposited under the same sputtering and in-situ $H_{suputtering}$ configuration. Under the illumination with the 405 nm laser, there is no modification in the shape of the hysteresis loop (nor in the exchange bias field) in either of the two samples (see Figure S3 of the Supporting information), suggesting no photostrictive effect in these cases. Figure 1d shows the dependence of $(ahv)^2$ (where $a$ denotes the absorption coefficient, and $hv$ is the photon energy) on the photo-energy, as well as the optical absorption as a function of wavelength (inset) for the (011)-oriented PMN-PZT and PMN-PT single crystals. In the visible light region, PMN-PZT crystal exhibits a sharp absorption near 400 nm, presenting an overall stronger absorption across the whole wavelength region compared to PMN-PT. In addition, the absorption wavelength in PMN-PZT is higher than for PMN-PT, indicating a smaller optical band gap. The optical band gap ($E_g$) was evaluated from the absorption coefficient ($\alpha$) by fitting the data according to the following relation:[43]

$$\alpha h v = \alpha_0 (E_g - hv)^n \qquad (1)$$

where $\alpha_0$ is a constant and $n$ is the exponent that determines the type of electronic transition causing the absorption. Here, $n$ is 1/2, as both PMN-PZT and PMN-PT crystals exhibit a direct optical band gap. By extrapolating the linear part of the curve, the direct band gap energy ($E_g$) of the PMN-PZT crystal ($E_{g,PMN-PZT}$) is estimated to be 3.03 eV, which is smaller than that of the PMN-PT (3.11 eV). This is in line with values previously reported in the literature.[26] Hence, the photostrictive effect is observed exclusively in the PMN-PZT single crystal, as its bandgap is lower than the photon energy of the irradiation laser, $E_{photo,405nm}$ = 3.06 eV. This photostrictive effect originates from the internal electric field generated by the bulk photovoltaic effect upon illumination with the visible light. This internal electric field affects the arrangement of the ferroelectric domains, possibly resulting in a net strain effect.

As light may potentially cause local heating of the sample, one could think that the observed changes in the $H_{EB}$ values may arise from a photothermally induced increase of temperature, eventually getting closer to the blocking temperature of the system, which, is around 520 K for 20 nm-thick IrMn.[44] However, for the used laser power, the maximum temperature increases in PMN-PZT is less than 1 K.[26] Such a limited temperature increase would have negligible effects on the magnetic properties of FeGa thin films. To corroborate that this little heating does not contribute to the observed modification of $H_{EB}$ in our samples, M-H loops have been measured at 300 K and 305 K, as shown in Figure S4. After increasing T by 5 K, no obvious changes are observed in the loops. Note that the absence of light-induced effects in the reference sample grown on Si reconfirms the negligible role of thermal heating in the observed $H_{EB}$ modulation, since Si experiences a much higher heating effect compared to PMN-PZT when exposed to light due to its larger light absorption.[45] Thus, the contribution of thermal heating arising from light excitation can be essentially ruled out in our case, leaving the photostrictive effect as the dominant mechanism responsible for the observed effects.

To quantify the magnitude of light-induced strain in our system, we used a strain gauge that measured the strain responses, $\Delta L / L$, along the [100] and [0−11] directions in a PMN-PZT single crystal under illumination. The corresponding results are presented in Figure 2a and 2b, respectively. Under illumination with the 405 nm laser, $\Delta L / L$ for [100] and [0−11] directions are approximately $8 \times 10^{-5}$ and $1 \times 10^{-4}$, respectively. The inset in Figure 2b shows a cyclic illumination experiment, in which the laser was switched on and off multiple times, demonstrating the repeatability and stability of the photostrictive effect in PMN-PZT. The obtained strain values are comparable to those in some other perovskite ferroelectrics, like PLZT ($1 \times 10^{-4}$)[46] and BiFeO$_3$ ($3 \times 10^{-5}$),[47] or semiconducting CdS ($7.5 \times 10^{-5}$).[48] The photostrictive effect is known to largely depend on the sample thickness and light intensity, and the photostrictive efficiency can be evaluated as: $\eta = h \frac{(\Delta L/L)}{I}$, where $h$ is the sample thickness along the direction of the illumination, and $I$ is the light intensity.[17] Using the 405 nm laser with a light intensity of 0.6 W cm$^{-2}$, the photostrictive efficiency is $-3.8 \times 10^{-12}$ m$^3$ W$^{-1}$ along the [100] and $-5.3 \times 10^{-12}$ m$^3$ W$^{-1}$ along the [0−11] orientation of the PMN-PZT crystal. It is worth noting that the photostrictive efficiency in both directions shows negative values, indicating that the induced in-plane stress effect is overall compressive in PMN-PZT under illumination.

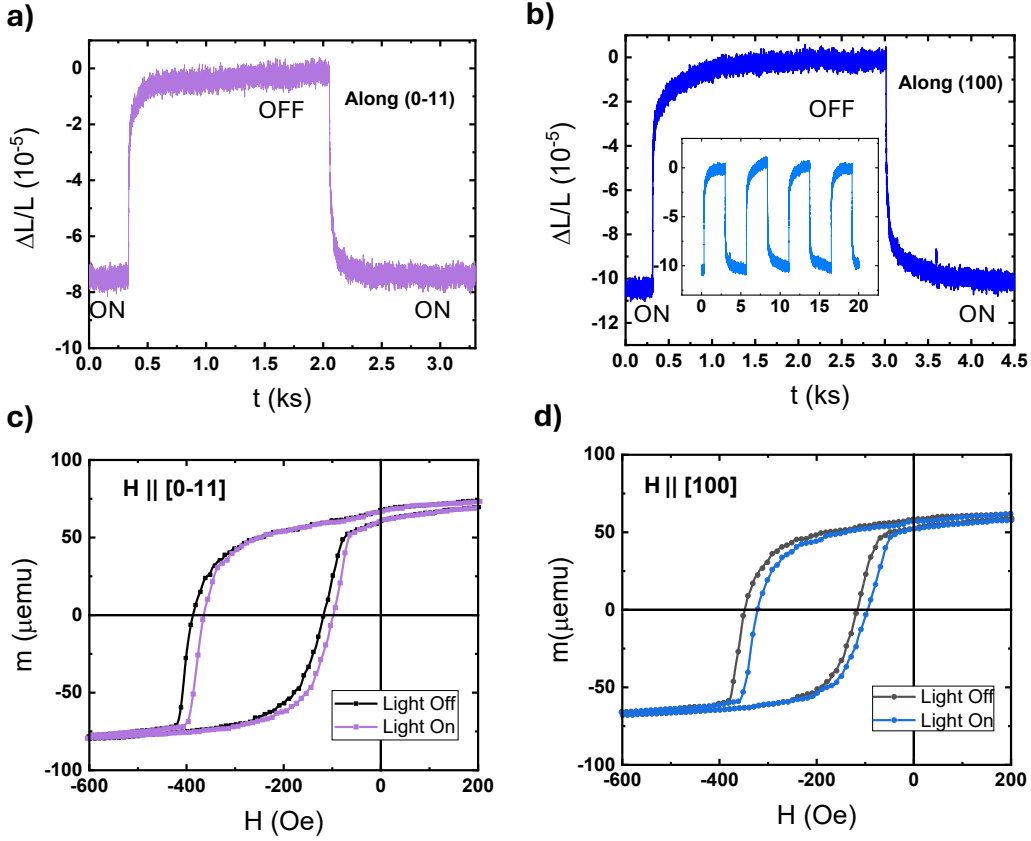

**Figure 2. a)** and **b)** Photostrictive performance under illumination in PMN-PZT single crystal along [0-11] and [100], respectively. The inset shows the photostrictive properties along [100] direction of PMN-PZT single crystal. **c)** and **d)** Magnetic hysteresis loops along the magnetic easy axis in dark and under illumination in $H_{sputtering}$ ∥ [0−11] and $H_{sputtering}$ ∥ [0−11] samples, respectively. Here the light intensity (0.36 W cm$^{-2}$) is the same for all the measurements.

The (001)-cut PMN-PZT single crystal is known to exhibit large and anisotropic transverse piezoelectric coefficients, which generate strong in-plane compressive stress along [100] (*i.e.*, $d_{32}$ = −1850 pC N$^{-1}$) and tensile stress along [0−11] (*i.e.*, $d_{31}$ = 599 pC N$^{-1}$) under the application of an external electric field [49]. Therefore, the rather isotropic compressive strain induced with light in PMN-PZT is not in agreement with the conventional electric-field induced magnetoelastic effect. Similarly, light-induced shape contraction was predicted along the polar direction of PLZT, but expansion is generally observed in PLZT film or bulk ceramics.[50] No anisotropy in the photostrictive effect has neither been mentioned for PLZT nor for BiFeO$_3$. This contradiction illustrates the complexity of the photostrictive mechanism in ferroelectrics, which cannot be simply attributed to the inverse piezoelectric effect. One possible explanation is that strain results from local deformation of ferroelectric domains driven by internal electric fields generated by the photovoltaic effect, rather than by full ferroelectric polarization switching under high applied voltages. The quantitative analysis of the unexpected isotropic strain induced with light in PMN-PZT is intriguing and requires further investigation.

The phenomenon describing the effect of light on magnetic properties was recently named the converse magneto-photostrictive (CMP) effect, and a coefficient to quantify this effect was then defined.[26] The CMP is analogous to the converse magnetoelectric effect (CME) and is also a function of the magnetic field. The converse magneto-photostrictive coupling coefficient was defined as follows:

$$\alpha_{CMP}^{\lambda}(H) = \mu_0 \partial M(H)/\partial I \quad (2)$$

expressed in s A$^{-1}$, where $\mu_0$ is vacuum permeability, $\partial M(H)$ is the variation of magnetization under a magnetic field $H$ and $\partial I$ is the change of light intensity at a certain wavelength $\lambda$. The CMP coupling coefficient is calculated by:

$$\alpha_{CMP}^{\lambda}(H) = \mu_0 \frac{M_{I=I_1}(H) - M_{I=I_0}(H)}{I_1 - I_0} \quad (3)$$

In this work, $\lambda$ = 405 nm, $I_1$ = 0.36 W cm$^{-2}$ and $I_0$ = 0. The relative change of magnetization is determined by subtracting the *M-H* hysteresis loops measured in dark and under illumination (see Figure 2c and 2d). Figure 3 presents the dependence of $\alpha_{CMP}^{405\,nm}(H)$ with the external magnetic field, $H$, for the two investigated samples, where the easy axis direction is aligned along the PMN-PZT [0−11] or the [100] directions, respectively. In both cases, starting at a positive field, $\alpha_{CMP}^{405\,nm}$ keeps at zero as the field decreases, until a sharp minimum is reached at the left coercive field $H_{C1}$, whereas another peak shows up at right coercive field, $H_{C2}$, when the magnetic field increases from negative to positive saturation. $\alpha_{ext}^{\downarrow}$ ( $\alpha_{ext}^{\uparrow}$ ) express the left (right) minima in $\alpha_{CMP}^{405\,nm}$ when $H$ decreases (increases). The converse magneto-photostrictive effect coefficient $\alpha_{CMP}^{405\,nm}(H)$ shows similar minima and magnitudes in both PMN-PZT crystal orientations (Figure 3a and b), with $\alpha_{ext}^{\downarrow}$ being always larger than $\alpha_{ext}^{\uparrow}$.

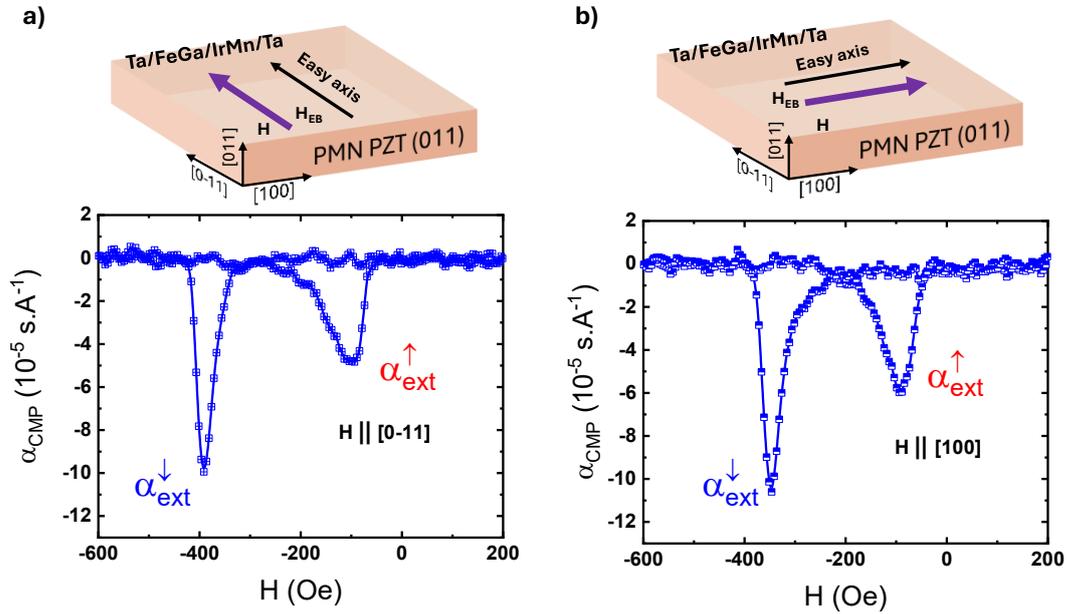

**Figure 3.** Magnetic field dependencies of converse magneto photostrictive coupling coefficients (CMP) of **a)** $H_{sputtering}$ ∥ [0−11] and **b)** $H_{sputtering}$ ∥ [100] in PMN-PZT/FeGa/IrMn heterostructure. $\alpha_{ext}^{\downarrow}$ and $\alpha_{ext}^{\uparrow}$ is the converse magneto photostrictive coupling coefficient extremum value as the magnetic field *H* decreases or increases, respectively.

Figure S5 shows magnetic hysteresis loops for PMN-PZT/FeGa/IrMn samples along different measuring angles, together with the angular dependences of $H_C$ and $H_{EB}$, both in the dark and under illumination, and both for $H_{sputtering}$ aligned along the [0−11] and the [100] crystallographic directions of the PMN-PZT. As expected, because of the magnetic easy axis direction dictated by $H_{sputtering}$, both $H_C$ and $M_R$ are higher for $\theta$ = 0°, and they progressively decrease with $\theta$, as the measuring angle changes from the easy to the hard axis directions of the films. The exchange bias field also decreases with $\theta$ and it vanishes at $\theta = 90°$, where the applied magnetic field is perpendicular to the exchange bias coupling direction. When the samples are illuminated with the laser, $H_C$ remains virtually unchanged, whereas $H_{EB}$ tends to systematically decrease. The reduction of $H_{EB}$ is particularly noticeable for $\theta = 0°$, where the light-induced change is around 21.4 Oe. To understand why illumination causes a decrease of $H_{EB}$, one needs to incorporate the "converse magnetostriction" or Villari effect[51] into the overall CMP phenomena. As previously described, when the sample is illuminated with the laser, a compressive photostrictive effect is induced in PMN-PZT both along [0−11] and [100], regardless of the direction of $H_{sputtering}$ (as shown in Figure 2a, b). In general, the magneto-elastic energy, $E_{mag\text{-}elast}$, arising from stress application in a polycrystalline ferromagnetic metal can be expressed as:[52]

$$E_{mag-elast} = -\frac{3}{2} \lambda_S \, \sigma \, cos^2(\varphi) \qquad (4)$$

where $\lambda_S$ refers to the material-dependent magnetostriction constant, $\sigma$ is the applied or induced stress, and $\varphi$ is the angle between the saturation magnetization and the stress direction. When $\lambda_S$ and $\sigma$ are both positive or both negative, the energy is minimized for $\varphi = 0°$ or 180° (no reorientation transition of the FM easy axis). Conversely, if $\lambda_S$ is positive but $\sigma$ is negative (or *vice versa*) then $E_{mag\text{-}elast}$ becomes a minimum for $\varphi = 90°$, meaning that the magnetic easy axis of the FM film tends to move away from its original direction). The room-temperature magnetostriction constant in FeGa polycrystalline alloys depends on the exact alloy composition but it attains positive values, typically around 200 ppm in bulk[53] and around 20 ppm in thin films.[54] Since the photo-induced in-plane stress in our system is always compressive (*i.e.*, negative), this means that the Villari effect tends to reorient the FeGa easy axis away from the initial $H_{sputtering}$ direction. In most exchange bias bilayer materials, $H_{EB}$ is maximum when the FM and AFM easy axes are parallel to each other, as in our as-prepared state samples (because both the FeGa and IrMn spin lattices will tend to align along $H_{sputtering}$ during growth of the films). When the samples are illuminated with the laser, there will be a tendency for the FeGa easy axis to deviate from its original direction, due to the positive $\lambda_S$ and the presence of compressive stress ($\sigma < 0$). Although the deviation is small and causes only a minor variation (< 5%) in the squareness ratio ($M_R/M_S$) of the loops when $\varphi$ is less than 45° (see Figure S5g, h), it is sufficient to induce a decrease in $H_{EB}$. Also, it has been reported that, in magnetostrictive FeGa films, the coercive field along the easy axis increases with the increase of the tensile stress and remains rather unaltered with increment of compressive stress,[55] which is in agreement with our experimental observations.

The strong light dependence of exchange bias observed in the studied FE/FM/AFM heterostructures holds promise for the development of novel devices capable of energy-efficient, remotely controlled wireless magnetization switching. Figure 4a illustrates the time evolution of magnetization under light illumination for the sample with $H_{sputtering}$ || [0−11], measured under an external magnetic field of –380 Oe (corresponding to $H_{C1}$). A one-minute waiting period was introduced prior to illumination to allow the magnetization to stabilize. Upon exposure to light, the magnetization reached saturation within two minutes and remained constant even after the light was turned off. A pronounced downward shift in magnetization is observed, indicating that the light-induced magnetization switching is deterministic and irreversible. Namely, removal of the light does not restore the magnetization to its original orientation. However, this issue can be addressed by applying a magnetic impulse, which reverses the magnetization and enables repeatable, reversible switching, as demonstrated in Figure 4b. Continuous magnetization reversals are achievable when a reset magnetic impulse of 1000 Oe is applied between each light exposure cycle. Similar results are observed for the $H_{sputtering}$ || [100] sample, (see in Figure S6, supporting information).

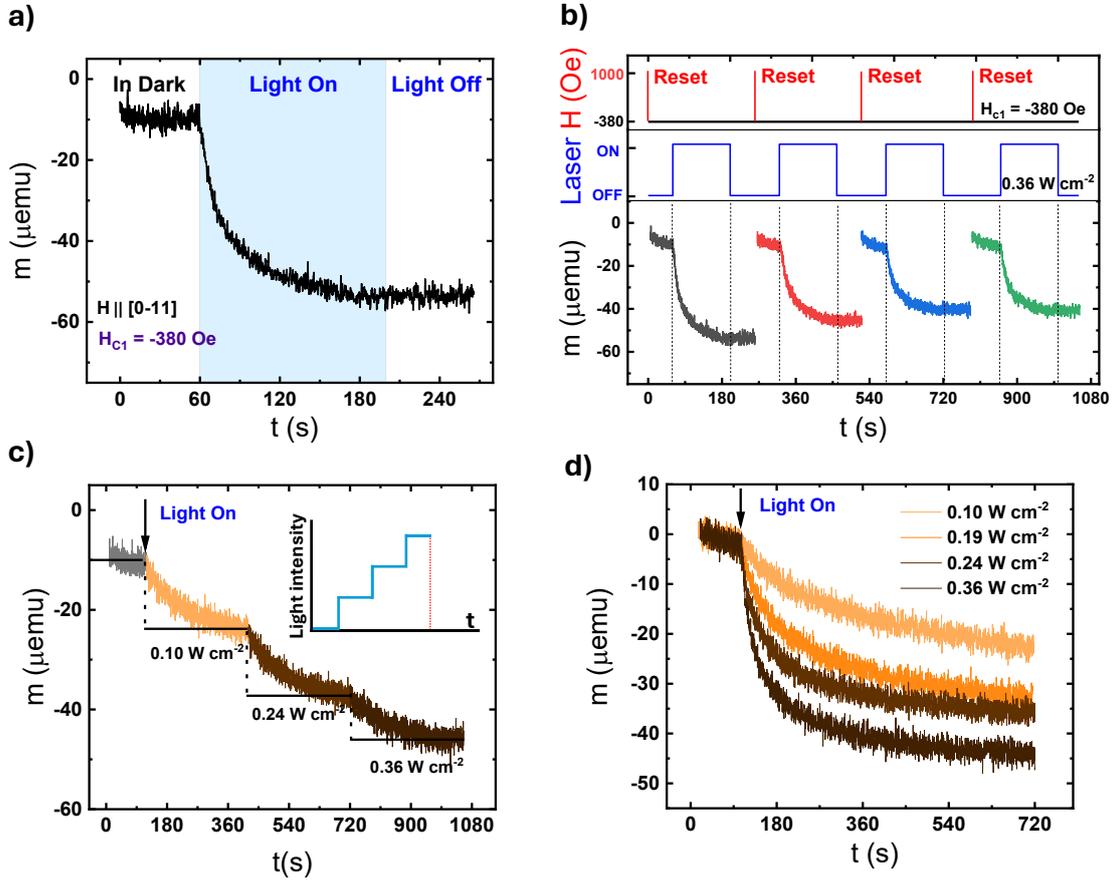

**Figure 4.** Light driven switching of magnetization though light modulating exchange bias field in PMN-PZT/FeGa/IrMn heterostructure with magnetic easy axis along [0−11], i.e. $H_{sputtering}$ || [0−11]. a) Magnetization switching by light at bias magnetic field of −380 Oe. b) Reversible magnetization switching by light with the assistance of magnetic field impulses. c) Magnetic multistate switching by varying the power intensity of the light as a function of time evolution. d) Magnetic states with different light power intensities, with an illumination duration of 10 minutes.

Next, we demonstrate the potential for analog control of magnetization using light, aiming toward the development of light-controlled multistate magnetic memory. As shown in Figure 4c, optically controlled multistate magnetization switching is achieved by varying the light intensity from 0.1 to 0.36 W cm$^{-2}$. With increasing light intensity, the magnetization reaches distinct intermediate levels, enabling multiple stable magnetic states. Furthermore, these states exhibit a superposition effect until magnetic saturation is reached. To evaluate the stability of this multistate switching, various light intensities were applied for over 10 min, as shown in Figure 4d. The saturation level of magnetization depends solely on the light intensity and does not increase with prolonged illumination. Under a fixed light intensity, the saturation value remains constant, while the time required to reach saturation decreases with increasing light power. This behaviour establishes a causal relationship between the optically induced internal electric field and the resulting strain. Importantly, the optical and magnetic inputs required for this process can be applied to the sample in a wireless manner (eventually using a permanent magnet positioned close to the sample instead of an electromagnet) and consume relatively low power compared to other memory devices relying on high electric currents to write and manipulate magnetic data. Thus, this light-intensity-controlled, magnetically assisted, multistate magnetization switching is fully repeatable and holds strong potential for future applications in optically driven magnetic memory devices.

## 3. Conclusion

We report significant optical control of exchange bias coupling at room temperature in a PMN-PZT/$Fe_{80}Ga_{20}$/$Ir_{20}Mn_{80}$ multiferroic heterostructure. Upon illumination, and with an external magnetic field applied parallel to the magnetic easy axis, either along the [100] or [0−11] crystallographic directions of the PMN-PZT single crystal, a measurable reduction in the exchange bias field is observed ($\Delta H_{EB}$ = 21.4 Oe). This effect arises from abnormal strain generated via the photostrictive response of the ferroelectric substrate, which induces compressive stress along both in-plane ferroelectric polarization directions. More importantly, we demonstrate a non-volatile light-induced modulation of exchange bias and magnetization switching, enabled by the photostrictive strain from the FE substrate and the inverse magnetostriction effect in FeGa. By combining light excitation with a low-power magnetic reset impulse, we achieve fully repeatable and energy-efficient magnetization switching. Additionally, we realize analog control of magnetization by tuning light intensity, which allows for optically controlled multistate magnetic memory. Multiple discrete magnetization levels are accessed by varying the light power, and these states remain stable over time, independent of the illumination duration. Compared with conventional current- or field-driven switching approaches, this light-modulated magnetization control significantly reduces energy consumption. These findings highlight the great potential of optically tuneable exchange bias and magnetization in multiferroic heterostructures for next-generation ultralow-energy, wireless, and multistate magnetic memory devices.

## 4. Materials and Methods

*4.1 Sample preparation*

The Ta(10 nm)/$Fe_{80}Ga_{20}$(5 nm)/$Ir_{20}Mn_{80}$(10 nm)/Ta(5 nm) heterostructure were deposited onto a (011)-cut Pb($Mg_{1/3}Nb_{2/3}$)$O_3$-Pb(Zr, Ti)$O_3$ (PMN-PZT) piezoelectric single crystal substrates using an AJA Internation magnetron system with a base pressure of 1 x $10^{-8}$ Torr at room temperature. PMN-PZT single crystals are commercially available as "CPSC160-95" from Ceracomp Co. Ltd., Korea. The $Fe_{80}Ga_{20}$ and $Ir_{20}Mn_{80}$ layers were DC sputtered from alloy targets with a 100 W power under Ar pressure of 3 x $10^{-3}$ Torr. During the depositions, a permanent magnet placed adjacent to the substrates, enabling to induce an exchange coupling between the ferromagnetic and antiferromagnetic bilayers either along [100] or [0−11] directions of PMN-PZT. For comparison purposes, identical multilayers were also grown on Si and Pb($Mg_{1/3}Nb_{2/3}$)$O_3$-$PbTiO_3$ (PMN-PT) substrates using the same deposition conditions.

*4.2 Magnetoelectric characterization*

All the magnetic hysteresis loops were measured using a vibrating sample magnetometer (VSM) from Lakeshore 8600 with a maximum applied field of 20 kOe. Hysteresis loops were recorded at different angles with respect to the in-situ magnetic field used during sputtering. Moment versus time curves were acquired under constant fields at room temperature, with and without light illumination. Argon atmosphere was introduced into the sample chamber while measuring loops slightly above room temperature (at 305 K).

To investigate the converse magneto-photostrictive effect, a 405 nm laser diode with maximum intensity 0.36 W cm$^{-2}$ was used to illuminate the sample from the backside of the substrate at a distance of 1 m. The incident angle of the laser beam was 90° (orthogonal to the substrate plane) for all the measurements.

*4.3 Ultraviolet–Visible spectroscopy (UV–Vis)*

Absorbance spectra were recorded in the 400−800 nm range using a JASCO V-780 spectrophotometer (Tokyo, Japan). Samples were mounted on a solid metal holder to ensure stable positioning and minimize stray light interference. All measurements were performed at room temperature, with baseline correction carried out using a clean metal holder as a reference.

*4.4 Strain measurements*

Strain measurements were measured using the resistance change of a strain gauge. A KFRB-02-120-C1-23 N15C2 Kyowa strain gauge was employed. A 410 nm laser with an intensity of 0.6 W cm$^{-2}$ was used to illuminate the substrate from the backside of the substrate. The incident angle of the laser beam is 90° (orthogonal to the substrate plane) for all the measurements.

*4.5 Structural/compositional characterization*

To investigate the structural and compositional characteristics of the samples, high-resolution transmission electron microscopy (HRTEM), high-angle annular dark-field scanning transmission electron microscopy (HAADF-STEM), and electron energy loss spectroscopy (EELS) were performed using a Spectra 300 STEM microscope (Thermo Fisher Scientific), operated at an accelerating voltage of 200 kV. These measurements were carried out at the Joint Electron Microscopy Center of the ALBA Synchrotron.

Cross-sectional lamellae were prepared by focused ion beam (FIB) milling after depositing a Pt-C protective layer to reduce charging effects and prevent oxidation or ion beam damage. The lamellae were then transferred onto Cu TEM grids for structural and compositional analysis.

**Acknowledgements**

Financial support by the European Research Council (2021-ERC-Advanced 'REMINDS' Grant Nº 101054687 and 2024-ERC-Proof of Concept 'SECURE-FLEXIMAG' Grant no. 101204328), the *Generalitat de Catalunya* (2021-SGR-00651), and the Spanish State Research Agency (PID2020-116844RB-C21 and TED2021-130453B-C22) is acknowledged. Views and opinions expressed are however those of the author(s) only and do not necessarily reflect those of the European Union or the European Research Council Executive Agency. Neither the European Union nor the granting authority can be held responsible for them.

**Conflict of interest**

The authors declare that they have no known competing financial interests or personal relationships that could have appeared to influence the work reported in this paper.

**Data and materials availability**

Data will be made available on request.

**Supporting Information**

**Light-modulated exchange bias in multiferroic heterostructures**

*Huan Tan[1,2]\*, Zheng Ma[1], Cynthia Bou Karroum[3], Matthieu Liparo[3], Jean-Philippe Jay[3], David Spenato[3], David T. Dekadjevi[3], Luís Martínez Armesto[1,2], Alberto Quintana[1,2], Jordi Sort[1,2,4]\**

[1]Departament de Física, Universitat Autònoma de Barcelona, E-08193 Cerdanyola del Vallès, Spain.

[2]Catalan Institute of Nanoscience and Nanotechnology (ICN2), CSIC and BIST, Campus UAB, Bellaterra, 08193 Barcelona, Spain

[3]Univ. Brest, Laboratoire d'Optique et de Magnétisme (OPTIMAG), UR 938, 29200 Brest, France.

[4]Institució Catalana de Recerca i Estudis Avançats (ICREA), Pg. Lluís Companys 23, E-08010 Barcelona, Spain

Keywords: Exchange bias, Photostrictive, Multiferroic

E -mail: Huan.Tan@uab.cat (H. Tan), Jordi.Sort@uab.cat (J. Sort)

**Figure S1.** a) TEM characterization of the multilayer structure, including individual EDX elemental maps (Fe, Ga, Ir, Mn, and Ta) and high-resolution TEM (HRTEM) image highlighting the polycrystalline nature of the layers. b) line profiles showing elemental distribution across the cross-section.

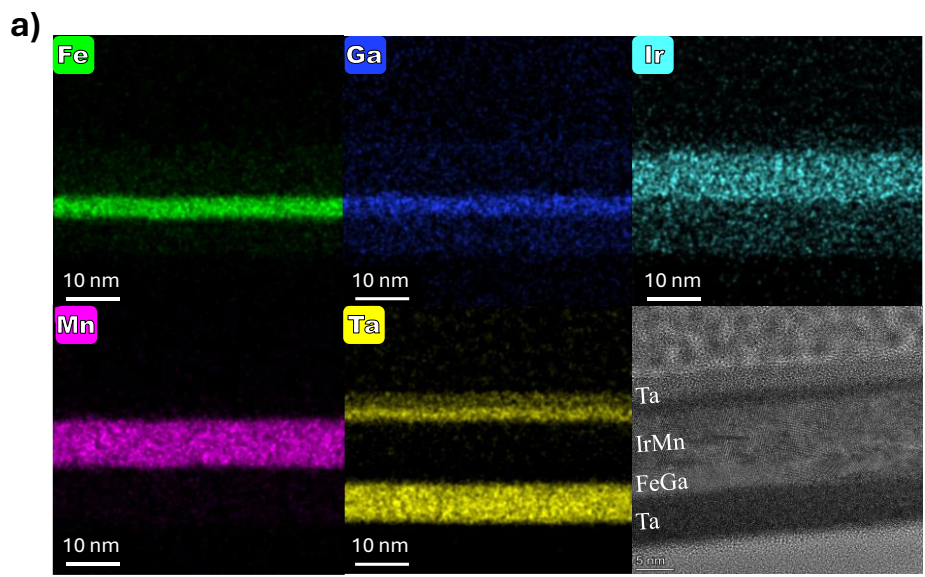

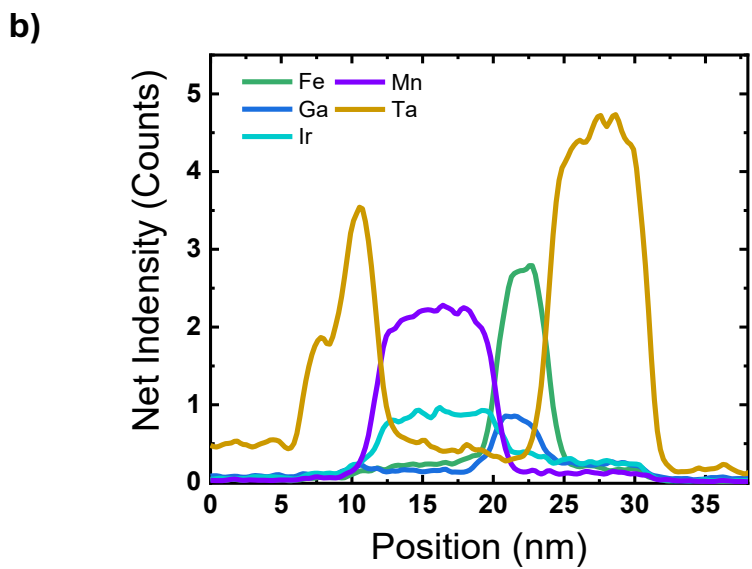

**Figure S2.** Successive magnetic hysteresis loops along the magnetic easy axis of **a)** $H_{sputtering} \parallel [0-11]$ and **b)** $H_{sputtering} \parallel [0-11]$ samples. **c)** and **d)** Variation of $-H_{EB}$ as a function of number of cycles ($n$) obtained from successive hysteresis loops of **a)** and **b)**. Note that "As-prepared" refers here to the samples after having been subjected to this successive hysteresis loops, before being illuminated with the laser.

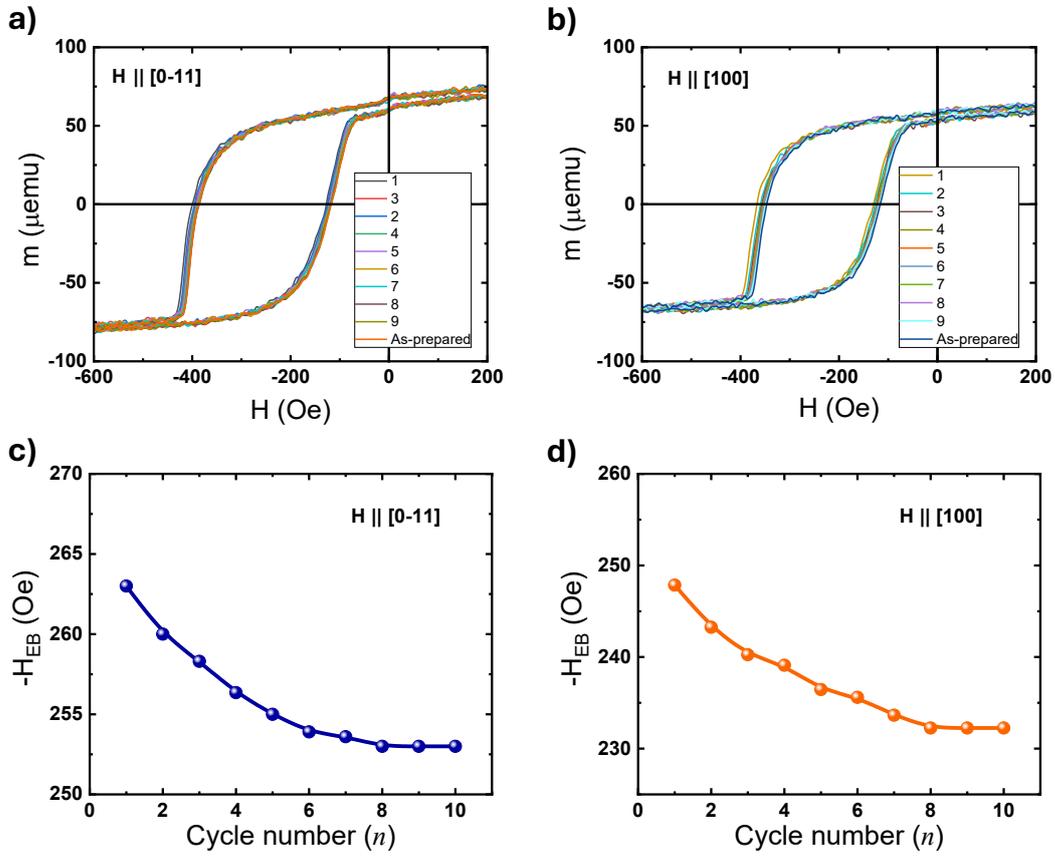

**Figure S3. a)** and **b)** Magnetic hysteresis loops along the magnetic easy axis sample in dark and under illumination in FM/AFM with Si and PMN-PT substrates, respectively.

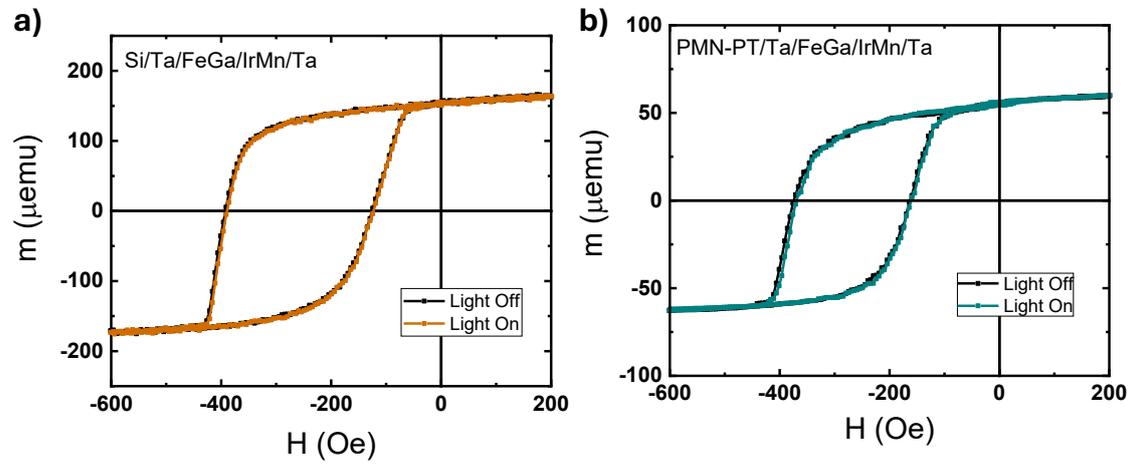

**Figure S4.** Magnetic hysteresis loops measured along the magnetic easy axis in $H_{sputtering}$ || [0−11] sample at room temperature ($T_R$) and after a temperature increase of 5 K ($T_H$).

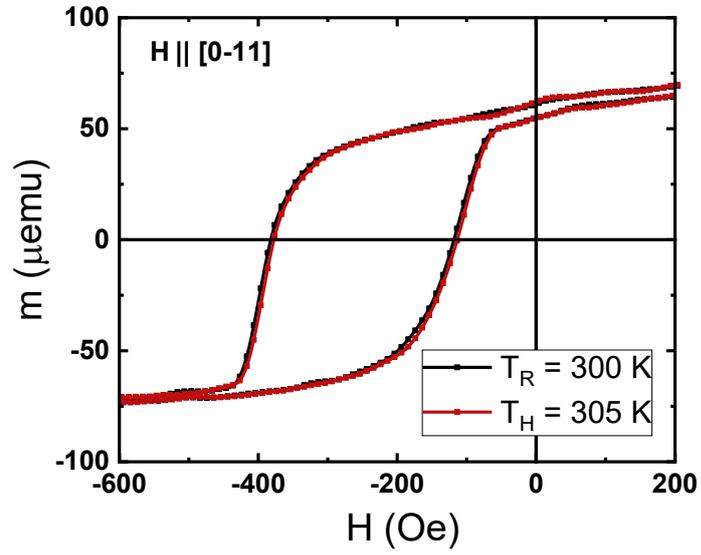

**Figure S5. a, c)** and **b, d)** Angular dependence (θ) of magnetic hysteresis loops for $H_{sputtering}$ || [0−11] and $H_{sputtering}$ || [100] in PMN-PZT/FeGa/IrMn heterostructure in dark and under illumination, respectively. **e)** and **f)** exchange bias field and coercive field values which were obtained from the hysteresis loops. **g)** and **h)** Squareness ($M_r/M_s$) of the hysteresis loop as a function of the angle θ. Here, θ represents the angle between the sputtering field direction (i.e., $H_{sputtering}$) and the magnetic field applied during the measurement of the hysteresis loops.

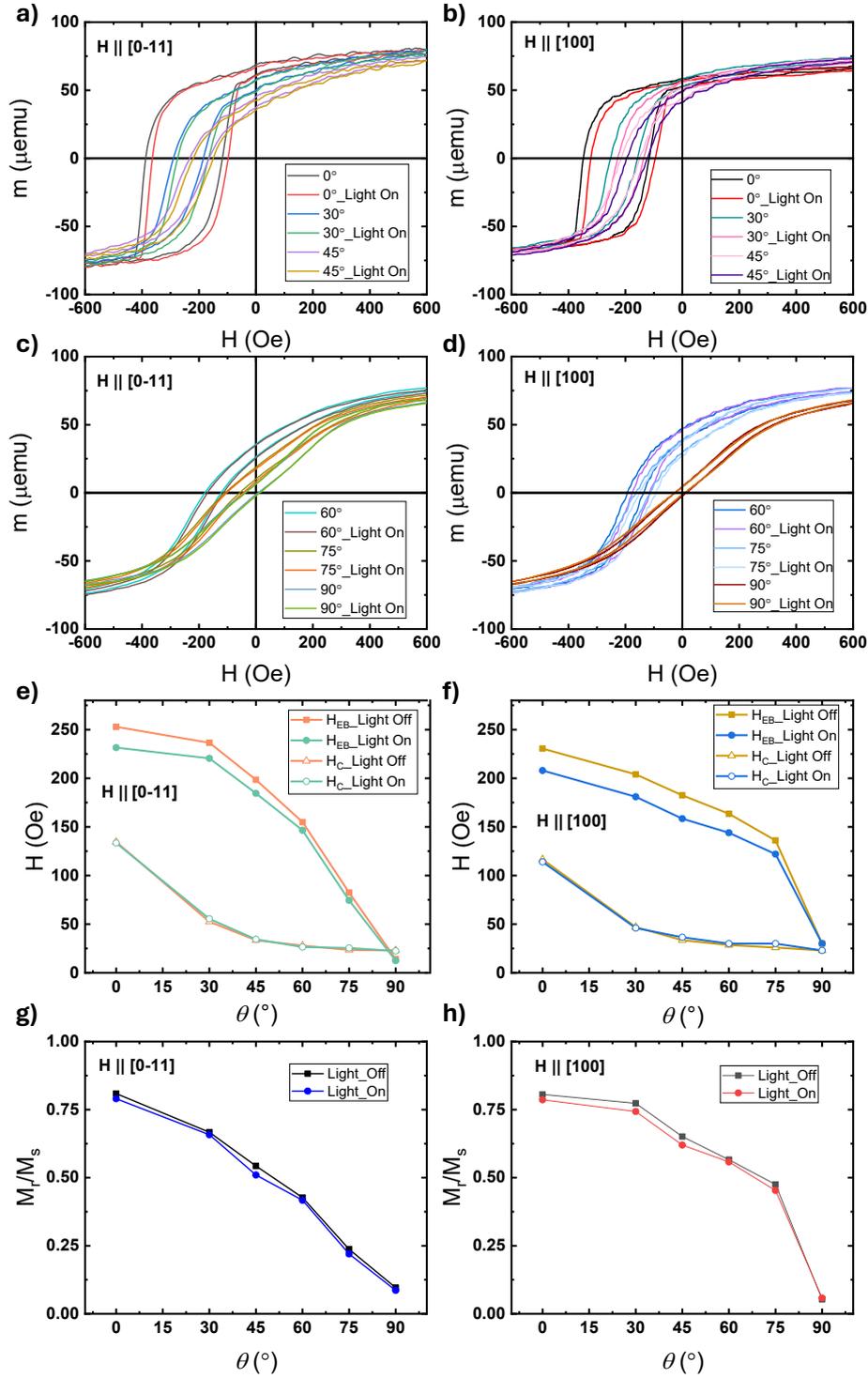

**Figure S6**. Light driven switching of magnetization though light modulating exchange bias field in PMN-PZT/FeGa/IrMn heterostructure with magnetic easy axis along [100], i.e. $H \parallel [100]$. a) Magnetization switching by light at bias magnetic field of −340 Oe. b) Reversable magnetization switching by light with the help of magnetic field impulses.

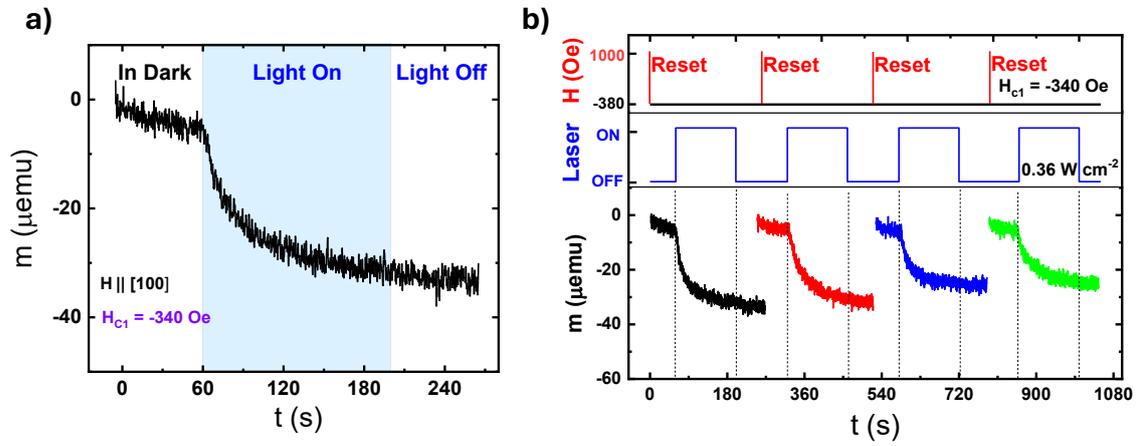